\documentclass[conference]{IEEEtran}
\IEEEoverridecommandlockouts

\makeatletter
\let\cl@chapter\undefined
\makeatletter

\usepackage[utf8]{inputenc}
\usepackage{amsmath}
\usepackage{amsthm}
\usepackage{hyperref}
\usepackage{cleveref}
\usepackage{amssymb}
\usepackage{siunitx}
\usepackage{txfonts}
\usepackage{cztsymb}
\usepackage{graphicx}
\usepackage{alltt}
\usepackage{xcolor}

\newcommand{\rat}{\mathbb{Q}}
\newcommand{\real}{\mathbb{R}}
\newcommand{\bool}{\mathbb{B}}

\renewcommand{\defs}{\triangleq}
\renewcommand\vec{\mathbf}

\newcommand{\qsem}[1]{\llbracket #1 \rrbracket_Q}
\newcommand{\qsub}{{\text{\!\tiny Q}}}
\newcommand{\scaleQ}{\mathop{*_\qsub}}
\newcommand{\qequiv}{\cong}
\newcommand{\qle}{\lesssim}

\newtheorem{definition}{Definition}
\newtheorem{theorem}{Theorem}

\title{Automated Reasoning for Physical Quantities, Units, and Measurements in Isabelle/HOL}
\author{
\IEEEauthorblockN{Simon Foster}
\IEEEauthorblockA{\textit{University of York}}
\and
\IEEEauthorblockN{Burkhart Wolff}
\IEEEauthorblockA{\textit{University Paris-Saclay}}
}

\hypersetup{
    colorlinks,
    linkcolor={red!50!black},
    citecolor={blue!50!black},
    urlcolor={blue!80!black}
}
\date{January 2023}

\begin{document}

\maketitle

\begin{abstract}
  Formal verification of cyber-physical and robotic systems requires that we can accurately model physical quantities that exist in the real-world. The use of explicit units in such quantities can allow a higher degree of rigour, since we can ensure compatibility of quantities in calculations. At the same time, improper use of units can be a barrier to safety and therefore it is highly desirable to have automated sanity checking in physical calculations. In this paper, we contribute a mechanisation of the International System of Quantities (ISQ) and the associated SI unit system in Isabelle/HOL. We show how Isabelle can be used to provide a type system for physical quantities, and automated proof support. Quantities are parameterised by \emph{dimension types}, which correspond to base vectors, and thus only quantities of the same dimension can be equated. Since the underlying ``algebra of quantities'' induces congruences on quantity and SI types, specific tactic support is developed to capture these. Our construction is validated by a test-set of known equivalences between both quantities and SI units. Moreover, the presented theory can be used for type-safe conversions between the SI system and others, like the British Imperial System (BIS).
\end{abstract}

\section{Introduction}
Cyber-physical systems use software to control interactions with the physical world, and so must account for quantifiable properties of physical phenomena. Modern physics uses quantities such as mass, length, time, and current, to characterise such physical properties. Such quantities are linked via an algebra to derived concepts such as velocity, force, and energy. The latter allows for a dimensional analysis of physical equations, which is the backbone of Newtonian Physics. In parallel, physicians developed their own research field called ``metrology'' for scientific study of the measurement of physical quantities. The integration of metrology into software engineering is of great importance to ensure that the physical components behave in a safe and predictable way~\cite{Burgueno2018RoboQuant,Hall2020Software-Quant,Flater2021-QuantSoftware}.

The international standard for quantities and measurements is distributed by the Bureau International des Poids et des Mesures (BIPM), which also provides the Vocabulaire International de M\'{e}trologie (VIM)~\cite{bipm-jcgm:2012:VIM}. The VIM actually defines two systems: the International System of Quantities (ISQ) and the International System of Units (SI, abbreviated from the French `Syst\`{e}me international d'unit\'{e}s'). The latter is also documented in the SI Brochure~\cite{SI-Brochure}, a standard that is updated periodically, most recently in 2019. Finally, the VIM defines concrete reference measurement procedures as well as a terminology for measurement errors.

Conceived as a refinement of the ISQ, the SI comprises a coherent system of units, built on seven base units, which are the metre, kilogram, second, ampere, kelvin, mole, candela, and a set of twenty-four prefixes to the unit names and unit symbols, such as milli- and kilo-, which may be used when specifying multiples and fractions of the units. The system also specifies names for 22 derived units, such as lumen and watt, for other common physical quantities. While there remains a wealth of measuring systems such as the British Imperial System (BIS), the SI can be considered as the de-facto reference behind them all.

The contribution of this paper is a mechanisation of the ISQ and SI in Isabelle/HOL, together with a deep integration into Isabelle's order-sorted polymorphic type system~\cite{Nipkow1991TypeClasses}. Our aim is two fold: (1) to provide a coherent mechanisation of the ISQ and its ontology of units; and (2) to support the use of units in formal specifications and models for cyber-physical systems~\cite{foster2020formal}. Our treatment of physical quantities allows the use of Isabelle's type system in checking for correct use of units. Since the algebra of quantities induces congruences on quantity types, specific tactic support is developed to capture these. Our construction is validated by a test-set of known equivalences between both quantities and SI units. Moreover, the presented theory can be used for type-safe conversions between the SI system and others, like the BIS.


Concretely, we introduce a novel parametric type for quantities, $N[\mathcal{D}, \mathcal{S}]$, where $N$ is the numeric type (e.g. $\rat$, $\real$), $\mathcal{D}$ is a dimension type (e.g. $\mathbf{L}$, $\mathbf{M}$, $\mathbf{T}$), and $\mathcal{S}$ is the system of units being employed. This accompanied by a formal ontology of units, which can be used in measurements. We can then write down specific quantities such as $20 \scaleQ metre :: \mathbb{R}[L, SI]$, which represents a measurement of 20 metres in the \textit{SI} (with dimension length), and $30 \scaleQ pound :: \mathbb{R}[M, BIS]$, which represents a measurement of 30 pounds in the \textit{BIS} (with dimension mass). Only quantities of the same dimension and unit system are comparable, and thus $20 \scaleQ metre = 30 \scaleQ pound$ is a type error. Nevertheless, we can convert between different unit systems, such that $metrify(30 \scaleQ pound) \approx 9.07 \scaleQ kilogram$. Our work employs several advanced features of Isabelle/HOL to implement the quantity type system without relying on the complexity of dependent types.


In summary, our contributions are:

\begin{enumerate}
\item an embedding of the ISQ into Isabelle/HOL, including dimensions, quantities, units, and conversions;
\item a sound-by-construction quantity type system for employs checking of dimensions and dimension coercions;
\item automated proof support for quantity conjectures;
\item a formal ontology of units from the VIM~\cite{bipm-jcgm:2012:VIM} and SI Brochure~\cite{SI-Brochure}, for use in specifications and models.
\end{enumerate}

The structure of our paper is as follows. In \S\ref{sec:related} we briefly survey related work to put our contributions into context. In \S\ref{sec:dimensions} we begin our contributions with our account of dimension types, which use a universe construction and type-class based characterisation. In \S\ref{sec:quantities} we use dimensions to implement the quantity type, $N[\mathcal{D}, \mathcal{S}]$, including automated proof. In \S\ref{sec:units}, we implement the SI unit system, and an associated ontology of units and equations. In \S\ref{sec:conversions} we describe conversions between unit systems, such as the \textit{SI}, \textit{CGS}, and \textit{BIS}. Finally, in \S\ref{sec:conclusions} we evaluate our work and conclude. Our entire theory development can be found on the Archive of Formal Proofs~\cite{Physical_Quantities-AFP}.


\section{Related Work}
\label{sec:related}
The need for physical quantities and measurement in software and formal specifications is widely acknowledged~\cite{HayesBrendan95,Burgueno2018RoboQuant,Hall2020Software-Quant,Flater2021-QuantSoftware}. Burgue\~{n}o et al.~\cite{Burgueno2018RoboQuant,Burgeno2019SoftwareQuant} argue the importance for safety of having physical quantities in robotic software models, and extend UML with types for quantities, dimensions, and units. Flater~\cite{Flater2021-QuantSoftware} argues for the extension of the SI standard with dimensions and units to support software metrology.

Quantity types are implemented in several mainstream numerical computation systems, such as MATLAB\footnote{\url{https://uk.mathworks.com/discovery/dimensional-analysis.html}} and Mathematica\footnote{\url{https://reference.wolfram.com/language/ref/Quantity.html}}, usually to support conversion between units, checking for unit consistency, and simplification of dimensions. Hall~\cite{Hall2020Software-Quant} describes a recent library for Python that implements quantities and facilitates dimensional analysis. Our work can serve as a baseline for verified implementations of the ISQ, particularly through the Isabelle code generator~\cite{Haftman2010-CodeGen}.

There have also been numerous direct implementations of ISQ and SI for programming languages.
Dimension Types have been presented by Kennedy~\cite{Kennedy94DimensionTypes,Kennedy09Units} for $F^\sharp$, as a way of parametrising data, and a more recent account along this line is by Garrigue and Ly~\cite{garrigue:hal-01503084}. These works directly implement a type system for dimensions and units in an ML-like language, while our approach formally derives such type inference inside the framework of parametric polymorphism and the framework of HOL. Thus, in contrast to direct implementations, our approach assures correctness by construction. 

Hayes et al.~\cite{HayesBrendan95} develop an extension of the Z specification language to incorporate units, dimensions, and quantities. Their main innovation is the addition of an operation $M \odot D$, which is effectively a type constructor for a quantity of numeric type $M$ and dimension type $D$. The latter has served as inspiration for our approach. However, their work lacks a supporting implementation, whereas we effectively provide a type system for quantities embedded into Isabelle/HOL. This is beyond the expressive power of Z. Moreover, unlike \cite{HayesBrendan95} our quantity types convey semantic information about the underlying dimensions, through our dimension universe construction, which can be used in reasoning.

Aragon~\cite{Aragon2004-SI} explores the algebraic structure of dimensions and quantities. He formalises quantities as ordered pairs called $q$-numbers, consisting of a complex number and a label, denoting the unit. He then explores the algebraic properties of unit labels and quantities. There is no explicit characterisation of dimensions, but units only. Nevertheless, the resulting properties have served as a benchmark for our work, for example showing that dimensions form an additive group.

Our work provides an implementation of the ISQ that is foundational, in that we precisely implement the quantity calculus, but also applicable, because it permits automatic checking of dimensions, efficient proof support, and code generation. We also provide a verified ontology of measurement units, which can be used in formal specifications and models~\cite{foster2020formal}. We are not aware of a comparable implementation of the ISQ in a proof assistant to date.

\section{Dimensions}
\label{sec:dimensions}
In this section we mechanise dimensions, which will be used in the following sections to parametrise physical quantities.

Dimensions are used to differentiate quantities of different kinds. For example, quantities of \SI{10}{kg} and \SI{10}{m} have the same magnitude, but are incomparable since they have the dimensions of mass and length, respectively. In the ISQ there are seven base quantities, including length, mass, and time, corresponding to seven base dimensions, which we will consider later in this article. The base dimensions are each denoted by a symbol, such as $L$, $M$, $T$ respectively, and a dimension is then a product of such symbols, each raised to an integer power. For example, the area quantity is represented by the dimension $L^2$, and the velocity quantity by $L \cdot T^{-1}$. Since we wish to support different unit systems, we here support a generic dimension system based on vectors.

\begin{figure*}
  \centering\includegraphics[width=.7\linewidth]{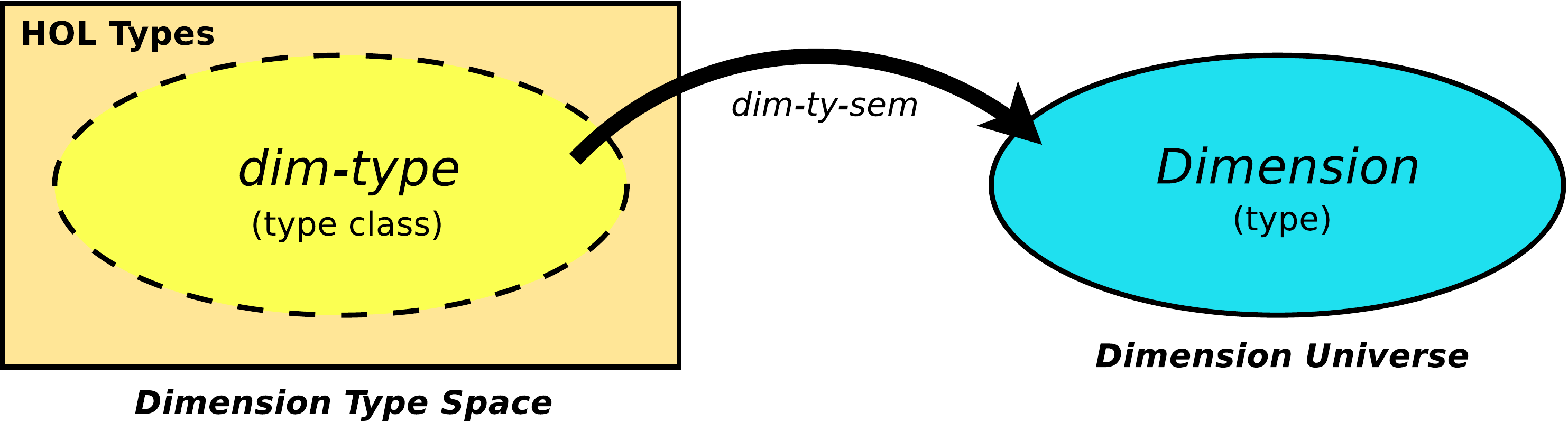}

  \caption{Mapping dimension types into the dimension universe}
  \label{fig:dim-mapping}
\end{figure*}

In a type theoretical context, dimensions can be seen as a parameter for physical quantities. Specifically, we can conceptually parametrise a quantity by its dimension, and since the equality and order relations are typically homogeneous, we can only compare quantities with the same dimension. However, to achieve this in a theorem prover like Isabelle/HOL, which lacks dependent types, we need to characterise dimension types using a different mechanism. In our case, we use type classes, which effectively allows us to isolate a given subset of type constructors which can be used to define dimensions. However, we first need to characterise a universe that these type constructors will be closed under. Consequently, we begin by defining a universe of dimensions, and then later use type classes to effectively define a homomorphism between dimension types and this universe.

This overall approach is illustrated in Figure~\ref{fig:dim-mapping}. We define (1) the dimension universe; (2) a type class (\textit{dim-type}) to syntactically characterise a class of types that characterise dimensions; (3) define a set of unitary types and type constructors that instantiate \textit{dim-type}, and can be used to parametrise quantities. Effectively, this achieves an inductively defined family of types over the dimension arithmetic operators. Our approach is generic, and can be applied to different measurement systems, though our focus is on the ISQ for the moment.

\subsection{Universe of Dimensions}

We begin by defining the dimension universe, the core operators for constructing dimensions, and their properties. If we assume there are $n \in \nat$ base quantities, then a dimension has the form $d_1^{\vec{x}_1} \cdot d_2^{\vec{x}_2} \cdots d_n^{\vec{x}_n}$, a product of dimension symbols ($d_i$) each raised to a power drawn from the vector $\vec{x}$. The encoding of the dimension vector $\vec{x}$ in Isabelle is shown below.

\begin{definition}[Dimension Vectors]
$$\textbf{typedef}~(N, I)~\textit{dimvec} = (\textit{UNIV} :: (I::enum) \Rightarrow N)$$
\end{definition}

\noindent The \textbf{typedef} command in Isabelle/HOL introduces a new type characterised by a non-empty subset of an existing type. In this case, we introduce a new type \textit{dimvec} with two parameters $N$ and $I$. We choose the set $UNIV$ of all total functions from $I$ to $N$ as the characteristic set. In future type definitions, we will use \textbf{typedef} to further restrict the subset. 

Conceptually, a dimension vector is simply a total function from an enumerable index type $I$, for the possible dimensions, to a numeric type $N$, which should minimally form a ring (e.g. $\num$). The enumerable ($enum$) sort constraint ($I :: enum$) requires that $I$ is isomorphic to a list of values, and is thus also a finite type. In the ISQ we have $I = \{\mathbf{L}, \mathbf{M}, \mathbf{T}, \mathbf{I}, \Theta, \mathbf{N}, \mathbf{J}\}$, for example.

Next, we define the core dimension constructors. For simplicity, we present these as functions using $\lambda$-terms, but they are technically defined using the lifting package~\cite{Huffman13Lifting}.

\begin{definition}[Dimension Constructors]
  $$\mathbf{1} \defs (\lambda i.~ 0) \quad \textbf{b}(i) \defs \mathbf{1}(i \mapsto 1)$$
  $$\vec{x} \cdot \vec{y} \defs (\lambda i.~ \vec{x}(i) + \vec{y}(i)) \quad \vec{x}^{-1} \defs (\lambda i.~ - \vec{x}(i))$$
\end{definition}
\noindent Here, $\mathbf{1}$ denotes a null dimension, which does not map to any physical quantity. It can characterise dimensionless quantities, such as mathematical constants ($\pi$, $e$, etc.) and functions.

The function $\textbf{b}(i)$, for $i \in I$, constructs a \emph{base dimension} from the base quantity $i$ by updating the mapping for $i$ in $\mathbf{1}$ to have the power $1$. A base dimension has exactly one entry in the vector mapping to 1, with the others all 0. We also define a predicate $\textit{is-BaseDim} :: (N, I)~\textit{dimvec} \Rightarrow \bool$, which determines whether a dimension vector corresponds to a base dimension.

A product of two dimensions ($\vec{x} \cdot \vec{y})$ simply pointwise sums together all of the powers, and an inverse ($\vec{x}^{-1}$) negates each of the powers. We can also now obtain division using the usual definition: $\vec{x} / \vec{y} \defs \vec{x} \cdot \vec{y}^{-1}$. With these definitions, we can prove the following group theorem:

\begin{theorem} If $(N, +, 0, -)$ forms an abelian group then also $((N, I)\textit{dimvec}, \cdot, \mathbf{1}, {}^{-1})$ forms an abelian group.
\end{theorem}
\noindent The abelian group laws can therefore be used to equationally rewrite dimension expressions, which is automated using Isabelle's simplifier.

Another avenue to efficient proof for dimensions is provided through the Isabelle code generator~\cite{Haftman2010-CodeGen}. Since the set of base quantities $I$ is enumerable, we can always convert a dimension vector to a list of $N$, and vice-versa. We achieve this using a function \textit{mk-dimvec}, which converts a list of $N$ with length $|\!I\!|$ to a dimension vector in $N$.

\begin{definition}[Converting Lists to Dimensions] $ $ %
\label{def:list-to-dim}

\vspace{1ex}
  \noindent
  $$
  \textit{mk-dvec}(ds) \defs \left(\begin{array}{l} \mathbf{if}~ (length(ds) = ~|\!I\!|) \\
                                       \mathbf{then}~ (\lambda d.~ ds(\textit{enum-ind}(d))) ~\mathbf{else}~ \mathbf{1}\end{array}\right)
  $$
\end{definition}
\noindent Since $I$ is enumerable, every dimension can be assigned a natural number, which also denotes its position in the underlying list. The function $\textit{enum-ind} :: (I\!::\!enum) \Rightarrow \nat$ extracts this positional index of a value in an enumerable type. For ISQ, we have $\textit{enum-ind}(\mathbf{L}) = 0$ and $\textit{enum-ind}(\mathbf{T}) = 2$, for example. We can then construct a dimension from a list $ds$ simply by looking up the value at the enumeration index.

Every possible dimension can be constructed using \textit{mk-dvec}, and so we can use it as a so-called ``code datatype'' for the Isabelle code generator. Dimensions are then encoded in SML or Haskell as an algebraic datatype with a single constructor corresponding to \textit{mk-dvec}, for example:

\begin{alltt}
 \textbf{datatype} ('a, 'b) dimvec = Mk_dimvec of 'a list
\end{alltt}

\noindent We then prove code equation theorems for the group operators, which are homomorphism laws, and enable efficient execution:

\begin{theorem} For a dimension vector space $(N, I)~dimvec$, with $|\!xs\!| \,=\, |\!ys\!| \,=\, |\!I\!|$, the following code equations hold:
\begin{align*}
  \mathbf{1} &= \textit{mk-dvec}\,(\textit{replicate}\,|\!I\!|\,0) \\
  \textit{mk-dvec}(xs) \cdot \textit{mk-dvec}(ys) &= map~(\lambda (x, y).\, x + y)~(zip~xs~ys) \\
  (\textit{mk-dvec}(xs))^n &= \textit{mk-dvec}~(map~(\lambda x.\, n \cdot x)~xs \\
  (\textit{mk-dvec}(xs))^{-1} &= \textit{mk-dvec}~(map~(\lambda x.\, - x)~xs
\end{align*}
\end{theorem}
\noindent These theorems give concrete definitional equations for the executable functions on the datatype. The null dimension is a list of $0$ powers of length $|\!I\!|$. Multiplication of two equilength lists $xs$ and $ys$ is pairwise addition of each element. Raising to the $n$th power multiplies each list element by $n$. Taking the inverse power negates each element. With such equations we can perform efficient dimension arithmetic on dimensions constructed from lists.

Dimensions in the ISQ are represented using the concrete dimension index type \textit{sdim}:

\begin{definition}[ISQ Base Quantities]
\label{def:isq-base-quant}
\begin{align*}
\textbf{datatype}~ sdim &= Length | Mass | Time | Current \\
                        &| Temperature | Amount | Intensity \\[1ex]
\textbf{type-synonym}~ Dimension &= (\num, sdim)~dimvec
\end{align*}
\end{definition}
\noindent
It suffices to show that \textit{sdim} is enumerable, using a type class instantiation, and then we can create a specific type synonym \textit{Dimension}, for dimension vectors in the ISQ. For convenience, we then define dimension vectors for each of the base quantities, for example $\mathbf{L} \defs \mathbf{b}(Length)$.

\subsection{Dimension Types}
\label{sec:dim-types}

Having defined our dimension universe, the next step is to characterise the family of dimension types. These dimension types will be used to parameterise our quantities, and ensure only quantities of the type dimension may be compared. We avoid the need for dependent types by first introducing a type class for dimension types.

\begin{definition}[Dimension Type Classes] $ $ %

\vspace{1ex}

$\begin{array}{l}
\textbf{class}~\textit{dim-type} = unitary~ + \\
~~ \textbf{fixes}~\textit{dim-ty-sem} :: \mathcal{D}~itself \Rightarrow Dimension \\[2ex]
\textbf{class}~\textit{basedim-type} = \textit{dim-type}~ + \\
~~ \textbf{assumes}~ \textit{is-BaseDim}\!: \textit{is-BaseDim}(QD(\mathcal{D}))
\end{array} 
$

\end{definition}
\noindent A type class characterises a family of types that each implement a given function signature with certain properties, such as algebraic structures like monoids and groups. The \textbf{class} command introduces a type class with a given name, potentially extending existing classes. The \textbf{fixes} subcommand declares a new typed symbol in the signature, and \textbf{assumes} introduces a property of the symbols in the signature.

The \textit{dim-type} class characterises a $unitary$ type $\mathcal{D}$ (i.e. a type with cardinality 1) and associates it to a particular dimension. The type $\mathcal{D}~itself$ represents a type as a value in Isabelle/HOL. Thus, \textit{dim-ty-sem} can be seen as a function from types inhabiting the \textit{dim-type} class to particular dimensions, as shown in Figure~\ref{fig:dim-mapping}. We can use the syntactic constructor $TYPE(\alpha)$ to obtain a value of type $\alpha~itself$, for a particular type $\alpha$. This effectively introduces an isomorphism between dimensions at the value level and the type level. For convenience, we introduce the notation $QD(\mathcal{D}) \defs \textit{dim-ty-sem}~\textit{TYPE}(\mathcal{D})$, which obtains the dimension of a given dimension type. The class \textit{basedim-type} further specialises \textit{dim-type} by requiring that the mapped dimension is a base dimension.

We use these classes to capture the set of type constructors for dimension types. First we construct types to denote the base dimensions, as unitary types. For example, we define the type length as below:
$$\textbf{typedef}~Length = (UNIV :: unit~set)$$
which exploits the fact that a type definition generates a fresh type name from a set
(in this case, the set that just contains the only element of the \textit{unit} type).
Though there is a seeming clash with the \textit{Length} constructor introduced in \cref{def:isq-base-quant}, these names inhabit different name spaces. \textit{Length} here is a ``tag type'' whose members do not convey information, but  represent dimension types syntactically.

We define seven such types, one for each of the ISQ base quantities, and also a further special type called \textbf{1}, which corresponds to a dimensionless quantity. Each of the base dimensions instantiates the \textit{basedim-type} class by mapping to the corresponding dimension symbol introduced in the previous section, such that, for example, $QD(Length) = Length$.

Next, we introduce the arithmetic operators for dimensions at the type level. The product and inverse type constructors are defined as shown below:
\begin{align*}
    &\textbf{typedef}~(\mathcal{D}_1\!::\!\textit{dim-ty},\mathcal{D}_2\!::\!\textit{dim-ty})~\textit{DimTimes} = (\textit{UNIV} :: unit set) \\
    &\textbf{typedef}~(\mathcal{D}\!::\!\textit{dim-ty})~DimInv = (\textit{UNIV} :: unit set)
\end{align*}
They are similarly tag types, but the parameters must inhabit the \textit{dim-ty} class. This ensures that the dimension types are closed under products and inverse. Using these type constructors, and the base dimension types, we can inductively define algebraic dimensions at the type level. We assign the type constructors the following implementations of \textit{dim-ty-sem}:
\begin{definition}[Semantic Interpretation of Dimension Types] \label{def:sem-dim} 
\begin{align*}
    \textit{dim-ty-sem}(d :: (\mathcal{D}_1, \mathcal{D}_2)~\textit{DimTimes}) &= QD(\mathcal{D}_1) \cdot QD(\mathcal{D}_2) \\
    \textit{dim-ty-sem}(d :: (\mathcal{D})~\textit{DimInv}) &= QD(\mathcal{D})^{-1}
\end{align*}
\end{definition}
These link together the type constructors and the underlying dimension operators. The semantics of a \textit{DimTimes} type calculates the underlying value-level dimension of each parameter $\mathcal{D}_1$ and $\mathcal{D}_2$, and multiplies them together. The \textit{DimInv} type similarly calculates the dimension and then takes the inverse. We give these type constructors the usual mathematical syntax, so that we can write dimension types like $\mathbf{M} \cdot \mathbf{L}$ and $\mathbf{T}^{-1}$. We also define a type synonym for division, namely $(\mathcal{D}_1, \mathcal{D}_2)~\textit{DimDiv} \defs \mathcal{D}_1 \cdot \mathcal{D}_2^{-1}$, and give it the usual syntax. Moreover, we define a fixed number of powers and inverse powers at the type level, such as $\mathcal{D}^{-3} = (\mathcal{D} \cdot \mathcal{D} \cdot \mathcal{D})^{-1}$.

\begin{figure}[t]
  \centering
  \includegraphics[width=.6\linewidth]{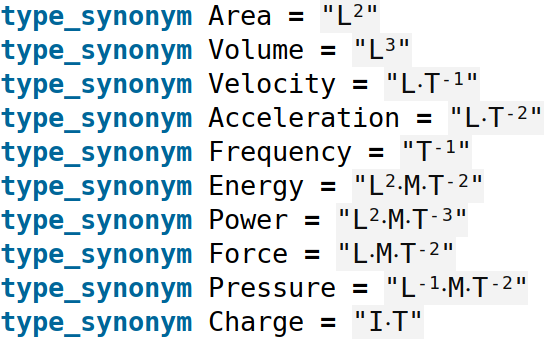}

  \caption{Derived dimension type expressions in Isabelle/HOL}
  \label{fig:derived-dimensions}

  \vspace{-1ex}
\end{figure}

We can now also create the set of derived dimensions specified in the ISQ using type synonyms. For example, we define $\textit{Velocity} \defs \mathbf{L} \cdot \mathbf{T}^{-1}$ and $\textit{Pressure} \defs \mathbf{L}^{-1} \cdot \mathbf{M} \cdot \mathbf{T}^{-1}$, which provides a terminology of dimensions for use in formal specifications. We show further example in Figure~\ref{fig:derived-dimensions}, which also demonstrates the mathematical syntax for dimensions implemented in Isabelle/HOL.

\subsection{Dimension Normalisation}
\label{sec:dim-norm}

Unlike dimensions at the value level, dimension types with different syntactic forms are incomparable, because they are distinct type expressions. For example, it is intuitively a fact that $\mathbf{L} \cdot \mathbf{T}^{-1} \cdot \mathbf{T} = \mathbf{L}$, which can be proved using the group laws. However, at the type level $\mathbf{L} \cdot \mathbf{T}^{-1} \cdot \mathbf{T}$ and $\mathbf{L}$ are different type expressions, and no built-in normalisation is available in Isabelle. As a result, we need to implement our own normalisation function, $\textit{normalise}(\mathcal{D})$, for ISQ dimensions in Isabelle/ML, so that quantities over dimensions with distinct syntactic forms can be related. Our normalisation function evaluates the dimension vector of a dimension expression, and then uses this to produce a normal form.

We implement dimension type evaluation using the ML function $\textit{typ-to-dim} :: typ \Rightarrow int~list$. It converts types formed of the base dimensions and dimension arithmetic operators into a dimension vector list, using the representation given in \cref{def:list-to-dim}. For example, $\textit{typ-to-dim}(\mathbf{L}) = [1, 0, 0, 0, 0, 0, 0]$ and $\textit{typ-to-dim}(\mathcal{D}^{-1}) = map~(\lambda x.\, - x)~(\textit{typ-to-dim}(x))$.

Having evaluated the dimension expression, we can use it to construct the normal form. This is an ordered dimension expression of the form $\mathbf{L}^{\vec{x}_1} \cdot \mathbf{M}^{\vec{x}_2} \cdot \mathbf{T}^{\vec{x}_3} \cdots \mathbf{J}^{\vec{x}_7}$, except that we omit terms where $\vec{x}_i = 0$. If every such term is $0$, then the function produces the dimensionless quantity, $\mathbf{1}$. As an example, $\textit{normalise}(\mathbf{T}^4 \cdot \mathbf{L}^{-2} \cdot \mathbf{M}^{-1} \cdot \mathbf{I}^2 \cdot \mathbf{M})$ yields the dimension type $\mathbf{L}^{-2} \cdot \mathbf{T}^4 \cdot \mathbf{I}^2$. This normalisation function is used later in this paper to facilitate coercion between quantities with distinct dimension expressions.

\section{Physical Quantities and Measurement}
\label{sec:quantities}
In this section we turn our attention to quantities themselves. As for dimensions, we will model quantities at both the value and type level. We also introduce the concept of \emph{measurement system}, which is used to specify the units being used for the different dimensions, such as metres for $\mathbf{L}$ and seconds for $\mathbf{T}$.

\subsection{Quantity Universe and Measurement Systems}
\label{sec:quant-univ}

We specify our quantity universe as a record with fields for the magnitude and dimension of the quantity.

\begin{definition}[Quantity Universe]
$$\begin{array}{rl}
  \textbf{record}~ (N, I\!::\!enum)\, Quantity =&\hspace{-1.5ex} mag\!::\!N\ \\ &\hspace{-1.5ex} dim\!::\!(int, I)\,dimvec
\end{array}$$
\end{definition}

\noindent The \textit{Quantity} type is parametric over a numeric type $N$ (e.g. $\rat$, $\real$), which should form a field, and the dimension index type $I$. The magnitude is then a number in $N$, and a dimension vector in over $I$. We can now specify the core arithmetic operators on quantities. For convenience of presentation, we use tuple syntax $(x, \mathcal{D})$, though in Isabelle the record fields are used.

\begin{definition}[Quantity Arithmetic Operators] \label{def:quant-arith}
  \begin{align*}
    0 &\defs (0, \mathbf{1}) \\
    1 &\defs (1, \mathbf{1}) \\
    (x, \mathcal{D}_1) \cdot (y, \mathcal{D}_2) &= (x \cdot y, \mathcal{D}_1 \cdot \mathcal{D}_2) \\
    (x, \mathcal{D})^{-1} &= (x^{-1}, \mathcal{D}^{-1}) \\
    (x, \mathcal{D}_1) / (y, \mathcal{D}_2) &= (x / y, \mathcal{D}_1 / \mathcal{D}_2) \\
    (x, \mathcal{D}) + (y, \mathcal{D}) &= (x + y, \mathcal{D}) \\
    (x, \mathcal{D}) - (y, \mathcal{D}) &= (x - y, \mathcal{D}) \\
    (x, \mathcal{D}_1) \le (y, \mathcal{D}_2) &\iff (x \le y \land \mathcal{D}_1 = \mathcal{D}_2)
  \end{align*}
\end{definition}

\noindent The arithmetic operators are overloaded in Isabelle/HOL, which is why they can validly appear on both sides of these equations. The ``0'' and ``1'' quantities are specified as dimensionless quantities with magnitude 0 and 1, respectively. Multiplication, inverse, and division are total operations that simply distribute through the pair. When multiplying two quantities, we need to multiply both the magnitudes and dimensions. For example, $(7, \mathbf{L}\cdot\mathbf{T}^{-1}) \cdot (2, \mathbf{T}) = (14, \mathbf{L})$. In contrast, addition and subtraction are partial operators that may be applied only when the two quantities have the same dimension. In Isabelle/HOL, the value of an addition or subtraction for different quantities of different dimensions is unspecified. Finally, the order on quantities is simply the order on the magnitudes, but with the requirement that the two dimensions are equal.

Quantities as formalised so far specify the form of dimension, but not the system of units being employed. For this, we extend the \textit{Quantity} type to create ``measurement systems'':

\begin{definition}[Measurement Systems]
$$
\begin{array}{l}
\textbf{record} (N, I\!::\!enum, \mathcal{S}\!::\!\textit{unit-system})~\textit{Measurement-System} \\
  \quad = (N, I)~Quantity + \textit{unit-sys} :: \mathcal{S}
\end{array}
$$  
\end{definition}

\noindent We extend the \textit{Quantity} record with an additional field \textit{unit-sys}. A measurement system is a quantity that specifies the system of units being used via an additional type parameter $\mathcal{S}$, which must inhabit the type class \textit{unit-system}. A unit system type is a unitary type that effectively allows us to tag quantities. This allows us to distinguish quantities using different systems of units, and so prevent improper mixing.

For example, the presence of the \textit{SI} tag means that a quantity of length is fundamentally measured in metres, whereas the presence of a tag such as \textit{BIS} may indicate that length is measured in yards. Later, we will use these to facilitate type-safe conversions between different unit systems.

All the arithmetic operators can be straightforwardly lifted to measurement systems. Since all such functions are monomorphic (e.g. of type $\alpha \Rightarrow \alpha \Rightarrow \alpha$), mixing of systems is avoided by construction.


\subsection{Dimension Typed Quantities}

Having defined our universe for quantities, we next enrich this representation with type-level dimensions. For expediency, we assume that all such quantities also have a measurement system attached. Moreover, we focus on quantities with dimensions from the ISQ.

\begin{definition}[Quantity Type]
$$
\begin{array}{l}
\textbf{typedef}~ (N, \mathcal{D}\!::\!\textit{dim-type}, \mathcal{S}::\textit{unit-system})~\textit{QuantT} \\ 
\quad = \{x :: (N, sdim, S)~\textit{Measurement-System}.~~ dim(x) = QD(\mathcal{D})\} 
\end{array}
$$
\end{definition}
\noindent The $(N, \mathcal{D}, \mathcal{S})~\textit{QuantT}$ type represents a quantity with numeric type $N$, dimension type $\mathcal{D}$, and unit system $\mathcal{S}$. The type definition introduces an invariant that requires that the dimension of the underlying quantity $x$ agrees with the one specified in the dimension type. At this level, we use \textit{sdim} as the concrete interpretation of dimensions, as this is required by the \textit{dim-type} class. For convenience, we introduce the type syntax $N[\mathcal{D}, \mathcal{S}]$ to stand for $(N, \mathcal{D}, \mathcal{S})~\textit{QuantT}$. Our \textbf{typedef} also induces two functions for converting between typed and untyped quantitities: $\textit{fromQ} :: N[\mathcal{D}, \mathcal{S}] \Rightarrow (N, \textit{sdim}, \mathcal{S})~\textit{Measurement-System}$ and \textit{toQ} in the opposite direction.

Lifting of arithmetic operators $x + y$ and $x - y$ is straightforward for typed quantities, since they are monomorphic and only defined when the dimensions of $x$ and $y$ agree. We can then easily show that typed quantities form an additive abelian group. We also define a scalar multiplication $\textit{scaleQ} :: N \Rightarrow N[\mathcal{D}, \mathcal{S}] \Rightarrow N[\mathcal{D}, \mathcal{S}]$, with notation $n \scaleQ x$, which scales a quantity by a given number without changing the dimension. We can then show that typed quantities form an additive abelian group, and a real vector space, with $(\scaleQ)$ as the scalar multiplication operator.

Things are more involved when dealing with general multiplication and division, since these need to perform dimension arithmetic at the type level. For example, if we have quantities $x :: \real[\mathbf{I}, SI]$ and $y :: \real[\mathbf{T}, SI]$, then multiplication of $x$ and $y$ is well-defined, and should have the type $\real[\mathbf{I}\cdot\mathbf{T}, SI]$. As a result, we introduce bespoke functions, $qtimes$, $qinverse$, and $qdivide$. We first give the types for these functions:
\begin{align*}
  qtimes &:: N[\mathcal{D}_1, \mathcal{S}] \Rightarrow N[\mathcal{D}_2, \mathcal{S}] \Rightarrow N[\mathcal{D}_1 \cdot \mathcal{D}_2, \mathcal{S}] \\
  qinverse &:: N[\mathcal{D}, \mathcal{S}] \Rightarrow N[\mathcal{D}^{-1}, \mathcal{S}] \\
  qdivide &:: N[\mathcal{D}_1, \mathcal{S}] \Rightarrow N[\mathcal{D}_2, \mathcal{S}] \Rightarrow N[\mathcal{D}_1 / \mathcal{D}_2, \mathcal{S}]
\end{align*}
\noindent The first function multiplies two quantities, with the same measurement system, and ``multiplies'' the dimension types using the type constructors introduced in \S\ref{sec:dim-types}. Technically, no multiplication computation takes place, but rather a type constructor denoting multiplication is inserted. Similarly, \textit{qinverse} represents the inverse of the parametrised dimension, and \textit{qdivide} stands for a division. What is achieved here is analogous to dependent types, though we require additional machinery for normalising dimension types (cf. \S\ref{sec:dim-norm}).

The definitions of \textit{qtimes} and \textit{qinverse} are obtained simply by lifting of the corresponding functions on quantities in Definition~\ref{def:quant-arith}, which is technically achieved using the \emph{lifting} package~\cite{Huffman13Lifting}. In order to do this we need to prove that the invariant of the \textit{QuantT} type is satisfied, which involves showing that the family of typed quantities is closed under the two functions. For \textit{qmult}, we need to prove that $dim(x \cdot y) = QD(\mathcal{D}_1 \cdot \mathcal{D}_2)$, whenever $dim(x) = QD(\mathcal{D}_1)$ and $dim(y) = QD(\mathcal{D}_2)$, which follows simply through Definitions \ref{def:sem-dim} and \ref{def:quant-arith}. For convenience, we give these functions the usual notation of $x \bullet y$, $x^{-1}$, and $x / y$, but in Isabelle we embolden the operators to syntactically distinguish them. With \textit{qtimes} and \textit{qinverse}, we can also define positive and negative powers, such as $x^{-2} = (x \bullet x)^{-1}$.

Equality ($x = y$) in HOL is a homogeneous function of type $\alpha \to \alpha \to \bool$; therefore, it cannot be used to compare objects of different types. Consequently, it cannot be used to compare quantities whose dimension types have different syntactic forms (e.g. $\mathbf{L} \cdot \mathbf{T}^{-1} \cdot \mathbf{T}$ and $\mathbf{L}$). This motivates a definition of \emph{heterogeneous (in)equality} for quantities:
\begin{align*}
  \textit{qequiv} &:: N[\mathcal{D}_1, \mathcal{S}] \Rightarrow N[\mathcal{D}_2, \mathcal{S}] \Rightarrow \bool \\
  \textit{qless-eq} &:: N[\mathcal{D}_1, \mathcal{S}] \Rightarrow N[\mathcal{D}_2, \mathcal{S}] \Rightarrow \bool
\end{align*}

\noindent These functions are defined simply by lifting the functions $(=)$ and $(\le)$ on the underlying quantities. They ignore the dimension types, but the underlying dimensions must nevertheless be the same as per the definitions in \S\ref{sec:quant-univ}. We give these functions the notation $x \cong y$ and $x \lesssim y$, respectively. Relation $(\cong)$ forms an equivalence relation, and $(\lesssim)$ forms a preorder. Moreover, $(\cong)$ is a congruence relation for $(\bullet)$, $({}^{-1})$, and $(\scaleQ)$.


\subsection{Proof Support}

We implement an interpretation-based proof strategy for typed quantity (in)equalities, which allows us to split a conjecture into two parts: (1) equality of the magnitudes; and (2) equivalence of the dimensions. This is supported by a function $\textit{magQ} :: N[\mathcal{D}, \mathcal{S}] \Rightarrow N$, with syntax $\qsem{-}$, which extracts the magnitude from a typed quantity. We can calculate magnitudes using interpretation laws, like the ones below:
  $$\qsem{x + y} = \qsem{x} + \qsem{y} \quad \qsem{x \bullet y} = \qsem{x} \cdot \qsem{y}$$
\noindent Such laws derive directly from the definition of the quantity operators in Definition~\ref{def:quant-arith}. The equation for addition implicitly makes use of the fact that $x$ and $y$ have the same dimension, and so addition is well-defined in the quantity universe. We then have the following transfer theorems, for the case of two quantitites with the same (syntactic) dimensions:
\begin{theorem}[Quantity Transfer Laws]
  $$x = y \iff (\qsem{x} = \qsem{y}) \qquad x \le y \iff (\qsem{x} \le \qsem{y})$$
\end{theorem}
\noindent In both cases, we need not check the equivalence of the dimensions as by construction we know that $x$ and $y$ have the same type, and so also have the same dimensions. It is sufficient simply to check the relation holds of the underlying magnitudes. For our heterogeneous (in)equality relations, we have the following transfer theorems:
\begin{theorem}[Heterogeneous Transfer Laws] Given quantities $x :: N[\mathcal{D}_1, \mathcal{S}]$ and $y :: N[\mathcal{D}_2, \mathcal{S}]$, we have
  $$x \cong y \iff \left(\qsem{x} = \qsem{y} \land QD(\mathcal{D}_1) = QD(\mathcal{D}_2)\right).$$ 
\end{theorem}
\noindent We can then prove heterogeneous equalities by calculation of the underlying magnitudes and dimensions, and use of the numeric and dimensions laws. We supply a proof method called \textit{si-simp}, which uses the simplifier to perform transfer and interpretation, and additionally invokes field simplification laws. An additional method called \textit{si-calc} also compiles dimension vectors (cf. Definition~\ref{def:list-to-dim}) using the code generator, and can thus efficiently prove dimension equalities. We can, for example, prove the following algebraic laws automatically:
\begin{theorem}[Quantity Algebraic Laws]
  \begin{align*}
    a \scaleQ (x + y) &= (a \scaleQ x) + (a \scaleQ y) \\
    x \bullet y &\qequiv y \bullet x \\
    (x \bullet y)^{-1} &\qequiv x^{-1} \bullet y^{-1}
  \end{align*}
\end{theorem}

\subsection{Coercion and Dimension Normalisation}

The need for heterogeneous quantity relations ($\qequiv$, $\qle$) can be avoided by the use of coercions to convert between two syntactic representations of the same dimension. Moreover, we can use Isabelle's sophisticated syntax and checking pipeline to normalise dimensions, and so automatically coerce quantities to a normal form. This improves the usability of the library, since the usual relations $(=)$ and $(\le)$ can be used directly.

We implement a function $\textit{dnorm} :: N[\mathcal{D}_1, \mathcal{S}] \Rightarrow N[\mathcal{D}_2, \mathcal{S}]$, which can convert between quantities with different dimension forms. In order to use it effectively, it is necessary to know the target dimension $\mathcal{D}_2$ in advance. It is defined below:
$$\textit{dnorm}(x) \defs (\mathbf{if}~QD(\mathcal{D}_1) = QD(\mathcal{D}_2)~\mathbf{then}~toQ~(\,fromQ(x))~\mathbf{else}~0)$$
The function checks whether the source and target dimensions ($\mathcal{D}_1$ and $\mathcal{D}_2$) are the same. If they agree, then it performs the coercion by erasing the types with \textit{fromQ} and reinstating the new dimension type with \textit{toQ}. Otherwise, it returns a valid quantity of the target dimension, but with magnitude $0$. For example, if we have $x :: \real[\mathbf{L} \cdot \mathbf{T}^{-1} \cdot \mathbf{T}, SI]$, then we can use $\textit{dnorm}(x) :: \real[\mathbf{L}, SI]$ to obtain a quantity with an equivalent dimension, since $QD(\mathbf{L} \cdot \mathbf{T}^{-1} \cdot \mathbf{T}) = QD(\mathbf{L})$. In general, for two equivalent quantities $x \qequiv y$, we have it that $\textit{dnorm}(x) = y$.

Next, we extend Isabelle's checking pipeline to allow dimension normalisation, so that $\mathcal{D}_2$ can be automatically calculated. We do this by implementing an SML function \textit{check-quant}, which takes a term and enriches it with dimension information. Whenever it encounters an instance of $\textit{dnorm}(t)$, it extracts the type of $t$, which should be $N[\mathcal{D}, \mathcal{S}]$. This being the case, we enrich the instance of \textit{dnorm} to have the type $N[\mathcal{D}, \mathcal{S}] \Rightarrow N[\textit{normalise}(\mathcal{D}), \mathcal{S}]$.

We then insert \textit{check-quant} into Isabelle's term checking pipeline. Technically, this is achieved using an Isabelle/ML API function called \texttt{Syntax\_Phases.term\_check}, which allow us to add a new phase into the term checking process. In this case, we add it after type inference has occurred so that we can use the unnormalised dimension type expression as an input to \textit{check-quant}.

The soundness of this transformation does not depend on the correctness of \textit{normalise}, since if an incorrect dimension is calculated, \textit{dnorm} will return 0. Nevertheless, the effect is to achieve something akin to dependent types, but in a first-order polymorphic type system.

\section{Unit Systems and the SI}
\label{sec:units}
In this section we implement units generally, and in particular the SI unit system. We then implement a formal ontology of derived units and prefixes, drawn from the VIM standard~\cite{bipm-jcgm:2012:VIM} and SI Brochure~\cite{SI-Brochure}. Our ontology consists of the 7 base units, 32 derived and accepted units, and 24 unit prefixes.

An SI unit is simply a quantity in the ISQ with magnitude 1, which is typically combined with a magnitude to describe a measurement. A \emph{base unit} for a particular unit system $\mathcal{S}$ is a quantity whose dimension is one of the base dimensions. Base units are described by the predicate $\textit{is-base-unit} :: N[\mathcal{D}, \mathcal{S}] \Rightarrow \bool$, defined as $\textit{is-base-unit}(x) \defs (\textit{mag}(x) = 1 \land \textit{is-BaseDim})$. We introduce the constructor $\textit{BUNIT}(\mathcal{D}, \mathcal{S})$, which constructs a base unit using the base dimension type $\mathcal{D}$ in the system $\mathcal{S}$, with $\qsem{\textit{BUNIT}(\mathcal{D}, \mathcal{S})} = 1$.

For the SI, we create a unitary type \textit{SI}, and instantiate the \textit{unit-system} class. We then define the 7 base units of the SI:

\begin{definition}[SI Base Units]
  $$\textit{metre} \defs \textit{BUNIT}(L, SI) \qquad \textit{kilogram} \defs \textit{BUNIT}(M, SI)$$
  $$\textit{ampere} \defs \textit{BUNIT}(I, SI) \qquad \textit{kelvin} \defs \textit{BUNIT}(\Theta, SI)$$
  $$\textit{mole} \defs \textit{BUNIT}(N, SI) \qquad \textit{candela} \defs \textit{BUNIT}(J, SI)$$
\end{definition}
\noindent
Since the \textit{second} is very often used as the unit of time, we characterise it as a polymorphic base unit, so that it can effectively exist in several systems. For convenience we create type synonyms, which allow us to specify units at the type level, for example $N~\textit{meter} \defs N[Length, SI]$, which is a quantity of dimension length in the SI.

We now at last have the facilitates to write quantities with SI units. At the basic level, we can write quantities like $20 \scaleQ metre$, which is the $metre$ unit scaled by 20, and has the the inferred type of $\real[L, SI]$. We can also write compound units, such as $10 \scaleQ (metre \bullet second^{-1})$, which has inferred type $\real[L \cdot T^{-1}]$. We can also prove unit equations like $(metre \bullet second^{-1}) \bullet second \qequiv metre$ using the \textit{si-calc} proof strategy, as shown below:

\begin{center}
  \includegraphics[width=.75\linewidth]{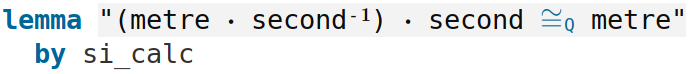}
\end{center}
\noindent Similarly we can use coercions to prove conjectures such as $\textit{dnorm}(((5 \scaleQ (metre / second)) \bullet (10 \scaleQ second)) = 50 \scaleQ metre$.

We can now turn our attention to constructing a formal ontology of derived SI units in Isabelle/HOL taken from the VIM and SI Brochure~\cite[page~137]{SI-Brochure}:

\begin{definition}[Core Derived Units]
  \begin{align*}
    hertz &\defs second^{-1} \\
    radian &\defs metre \bullet metre^{-1} \\
    steradian &\defs metre^2 \bullet metre^{-2} \\
    joule &\defs kilogram \bullet metre^{2} \bullet second^{-2} \\
    watt &\defs kilogram \bullet metre^{2} \bullet second^{-3} \\
    coulomb &\defs ampere \bullet second \\
    lumen &\defs candela \bullet steradian
  \end{align*}
\end{definition}

\noindent Isabelle can infer the dimension type of each such unit, for example \textit{watt} has the dimension $\mathbf{M} \cdot \mathbf{L}^2 \cdot \mathbf{T}^{-3}$. Radians and steradians have the dimensions $\mathbf{L} \cdot \mathbf{L}^{-1}$ and $\mathbf{L}^2 \cdot \mathbf{L}^{-2}$, which are distinct dimension types, but both semantically equal to the dimensionless quantity $\mathbf{1}$. Interestingly, it has been argued elsewhere than there should be a separate \textit{angle} dimension~\cite{Kalinin2019AnglesSI,Flater2021-QuantSoftware}, which would be necessary to formally distinguish them. Nevertheless, we choose to implement the SI as it is defined, though future extension is possible.

The SI defines 24 prefixes, which can be used to scale SI units. We give a selection of these below:

\begin{definition}[SI Prefixes]
   $$hecto \defs 10^2 \quad kilo \defs 10^3 \quad mega \defs 10^6 \quad giga \defs 10^9$$
   $$deci \defs 10^{-1} \quad centi \defs 10^{-2} \quad milli \defs 10^{-3} \quad micro \defs 10^{-6}$$
\end{definition}

\noindent Prefixes are not quantities, but simply abstract numbers in $N$, which can be used to scale units. For example, we can write a quantity such as $40 \scaleQ milli \scaleQ metre$.

The SI also has a notion of ``accepted'' units~\cite[page~145]{SI-Brochure}, which are quantities often used as units, but not technically SI because they have a magnitude other than 1. We give a selection of these below:

\begin{definition}[Accepted Non-SI Units]
  $$minute \defs 60 \scaleQ second \quad hour \defs 60 \scaleQ minute$$
  $$day \defs 24 \scaleQ hour \quad degree \defs (\pi / 180) \scaleQ radian$$
  $$litre \defs 1/1000 \scaleQ metre^3 \quad tonne = 10^3 \scaleQ kilogram$$
\end{definition}

\noindent These quantities can readily be treated as units in our mechanisation, though the type of such a quantity does not reflect the unit. For example, the units \textit{day}, \textit{hour}, and \textit{year} all have the dimension $\mathbf{T}$, as expected, meaning they are comparable. We can therefore prove unit equation theorems such as $1 \scaleQ hour = 3600 \scaleQ second$, $1 \scaleQ day = 86400 \scaleQ second$, and $1 \scaleQ hectare = 1 \scaleQ (hecto \scaleQ metre)^2$ using the \textit{si-simp} method, which can act as the basis for unit conversions. Similarly, we can use prefixes to express relations between derived quantities, such as $25 \scaleQ metre / second = 90 \scaleQ (kilo \scaleQ metre) / hour$.

The SI units are defined in terms of exact values for 7 physical constants~\cite[page~127]{SI-Brochure}. We define these in Isabelle:

\begin{definition}[Defining Constants of the SI]
  \begin{align*}
    \Delta v_{Cs} & = 9192631770 \scaleQ hertz \\
    \mathbf{c} &= 299792458 \scaleQ (metre \bullet second^{-1}) \\
    \mathbf{h} &= (6.62607015 \cdot 10^{-34}) \scaleQ (joule \bullet second) \\
    \mathbf{e} &= (1.602176634 \cdot 10^{-19}) \scaleQ coulomb \\
    \mathbf{k} &= (1.380649 \cdot 10^{-23}) \scaleQ (joule / kelvin) \\
    N_A &= 6.02214076\cdot 10^{23} \scaleQ (mole^{-1}) \\
    K_{cd} &= 683 \scaleQ (lumen/watt)
  \end{align*}
\end{definition}

\noindent $\Delta v_{Cs}$ is the hyperfine transition frequency of the caesium 133 atom. Constant $\mathbf{c}$ is the speed of light in a vacuum, and $\mathbf{h}$ is the Planck constant. Constant $\mathbf{e}$ is the elementary charge, and $\mathbf{k}$ is the Boltzmann constant. $N_A$ is the Avagadro constant. $K_{cd}$ is the luminous efficacy of monochromatic radiation of frequency $540 \cdot 10^{12}\, Hz$. These physical constants serve to ground measurements using a particular SI unit. With these constants, we can arrange their definitional equations to verify defining theorems for each unit, as shown below:

\begin{theorem}[Foundational Equalities]
  \begin{align*}
  second &\qequiv (9192631770 \scaleQ \textbf{1}) / \Delta v_{Cs} \\
  metre &\qequiv (\mathbf{c} / (299792458 \scaleQ \textbf{1})) \bullet second \\
  kilogram &\qequiv (\mathbf{h} / (6.62607015 \cdot 10^{-34}) \scaleQ \mathbf{1}) \bullet metre^{-2} \bullet second
  \end{align*}
\end{theorem}

\noindent The $second$ is equal to the duration of $9192631770$ periods of the radiation of the ${}^{133}Cs$ atom. The $metre$ is the length travelled by light in a period of $1/299792458$ seconds. For $kilogram$, the equation effectively defines the unit $\SI{}{kg.m.s^{-1}}$, and then applies the unit $\SI{}{m^{-2}.s}$ to obtain a quantity of dimension $\mathbf{M}$. Each equation is proved using \textit{si-calc}, which serves to validate our implementation of the SI. Finally, we complete our ontology of derived units~\cite[page~137]{SI-Brochure}:

\begin{definition}[Further Derived Units (selection)]
  \begin{align*}
    \textit{newton} &\defs kilogram \bullet metre \bullet second^{-2} \\
    \textit{pascal} &\defs kilogram \bullet metre^{-1} \bullet second^{-2} \\
    \textit{volt} &\defs kilogram \bullet metre^{2} \bullet second^{-3} \bullet ampere^{-1} \\
    \textit{farad} &\defs kilogram \bullet metre^{-2} \bullet second^{4} \bullet ampere^{2} \\
    \textit{ohm} &\defs kilogram \bullet metre^{2} \bullet second^{-3} \bullet ampere^{-2}
  \end{align*}
\end{definition}

\noindent Also, temperature in the SI is defined in Kelvin, but it is more usual to express temperature in terms of degree celcius. We therefore define $T\,{}^{\circ}\mathbf{C} \defs (T + 273.15) \scaleQ kelvin$, where $273.15$ is the freezing point of water. We can prove the corresponding unit equations, which show equivalences between SI units:

\begin{theorem}[Derived Unit Equivalences]
  $$joule \qequiv newton \bullet metre \quad watt \qequiv joule / second$$
  $$volt = watt / ampere \quad farad \qequiv coloumb / volt$$
\end{theorem}

\noindent The remaining derived units from the standard are all mechanised in Isabelle. Finally, as an application of our approach, we implement a selection of astronomical units:

\begin{definition}[Astronomical Units]
\begin{align*}
  \textit{julian-year} &\defs 365.25 \scaleQ day \\
  \textit{light-year} &\defs \textit{dnorm}(\mathbf{c} \bullet \textit{julian-year}) \\
  \textit{astromonical-unit} &\defs 149597870700 \scaleQ metre \\
  \textit{parsec} &\defs 648000 / \pi \scaleQ \textit{astronomical-unit}
\end{align*}
\end{definition}

\noindent The light year, astronomical unit, and parsec are all quantities of dimension $\mathbf{L}$. The light year is the distance travelled by light in one Julian year. We define it by multipliying $\mathbf{c}$ by \textit{julian-year} and normalising the result. The astronomical unit is the approximate distance between the earth and the sun. The parsec is the distance at which 1 astronomical unit subtends an angle of one arcsecond. We can give the parsec an exact mathematical value using Isabelle's Cauchy real characterisation of $\pi$.

\section{Unit Conversions and Non-SI Systems}
\label{sec:conversions}
In this section we describe unit conversion schemas, which can be used to convert quantities between different unit systems. Aside from the SI, other units systems remain in wide spread use today, notably other metric systems such as CGS (centimetre-gram-second), and imperial systems, including the United States Customary system (USC) and the British Imperial System (BIS). Interoperability with these systems therefore remains important. With our present system of quantities, we can already describe imperial units, in terms of the SI units, as shown below: 

\begin{definition}[Imperial Units in the SI]
  $$yard \defs 0.9144 \scaleQ metre \quad mile \defs 1760 \scaleQ yard$$
  $$pound \defs 0.4535937 \scaleQ kilogram \quad stone \defs 14 \scaleQ pound$$
  $$pint \defs 0.56826125 \scaleQ litre \quad gallon \defs 8 \scaleQ pint$$
\end{definition}

\noindent Here, we define the international yard and pound, units of length and mass, which were given exact metric definitions in 1959. From these, we derive units like the $mile$ and $stone$. The $pint$ is according to the imperial definition standardies in the UK in 1995, and similarly for the $gallon$. Such units can then be used to construct quantities in the usual way. However, this masks an inherent problem with units like yards, pounds, and pints: they have several definitions depending on the context.

Whilst the international yard is 0.9144 metres, the BIS yard has a slighty different definition of around 0.9143992 metres. This definition is based on a measurement of the imperial standard yard, a physical measure that was manufactured in 1845 and, after rigorous testing, made the official standard in 1855 by Act of Parliament~\cite{Bigg1963BritishYard,Zupko90Measurement}. The standard yard was then measured in 1895 against the metric standard, and found to have a length of 0.9143992 metres\footnote{Likely this discrepancy is due to the fact that the yard standard was slowly shrinking over time, a critical fact that was discovered later.}.

On the other hand, the USC has a slightly different definition again of around 0.9144018 metres, standardised by the 1866 Metric Law~\cite{NIST-SI}. Moreover, the volume unit ``gallon'' in the BIS and UCS have quite different definitions of 277.421 and 231 cubic inches (the 1707 ``wine gallon''~\cite{Zupko90Measurement}), respectively, and similarly for derived units, such as the pint. These inconsistencies are particularly a problem with historical measurements, which are more likely to use one of the older definitions.

Consequently, when precise measurements are crucial, it is necessary to characterise explicitly the unit system being employed, and define conversion factors between different systems. Even for metric systems, it is sometimes desirable to use different units, such as in the CGS system, where centimetres and grams are used as base units. In this case, we would also like the type system to enforce compatibility between measurements. We therefore formalise both unit systems and conversion schemas.

A conversion schema is a 7-tuple of rational numbers each greater than zero. Each rational number encodes a conversion factor for each of the dimensions of the ISQ. We define a type for conversion schemata, $\mathcal{S}_1 \Rightarrow_U \mathcal{S}_2$, which can be used to convert quantities between unit systems $\mathcal{S}_1$ and $\mathcal{S}_2$. Technically, we implement conversion schemata using a record type and type definition in Isabelle/HOL, whose definition is omitted for space reasons.

We define the identity conversion schema $id_C :: \mathcal{S} \Rightarrow \mathcal{S}$, which has $1$ for each of the factors. We can compose two conversion schemas $C_1 :: \mathcal{S}_1 \Rightarrow_U \mathcal{S}_2$ and $C_2 :: \mathcal{S}_2 \Rightarrow_U \mathcal{S}_3$ using the operator $C_2 \circ C_1 :: \mathcal{S}_1 \Rightarrow_U \mathcal{S}_3$, which pairwise multiplies each of the conversion factors in $C_1$ and $C_2$ respectively. Similary we can invert $C_1$ using $inv_C(C_1) :: \mathcal{S}_2 \Rightarrow_U \mathcal{S}_1$, which takes the reciprocal of each conversion factor. These operators induce a simple category of conversion schemas.

We use conversion schemas to define a quantity conversion function $\textit{qconv} :: (\mathcal{S}_1 \Rightarrow_U \mathcal{S}_2) \Rightarrow N[\mathcal{D}, \mathcal{S}_1] \Rightarrow N[\mathcal{D}, \mathcal{S}_2]$, whose definition is below:

\begin{definition}[Quantity Conversion]
  $$\textit{qconv}_C(m, \vec{d}) = \left(\left(\prod_{1 \le i \le 7}\, C_i^{\vec{d}_i}\right) \cdot m, \vec{d}\right)$$
\end{definition}
\noindent Given a quantity $(m, \vec{d})$, and a conversion schema $C$, the \textit{qconv} function calculates the conversion factor for the magnitude $m$ by raising each element of $C$ to the corresponding dimension element $\vec{d}_i$. For example, if we wish to convert cubic (international) yards to cubic metres, then we first need the conversion factor from yards to metres, which is 0.9144. Then, we take this value and raise it to the power of 3, and so the overall conversion factor is 0.764555. The dimension itself is unchanged by this operation, as expected.

The BIS is a non-metric standard for weights and measures in the UK, that was passed by an act of the UK parliament in 1824. It specifies the standard units for length and mass as the yard and pound, respectively. We model the BIS by creation of a unit system with the type \textit{BIS}, and define $\textit{yard} \defs BUNIT(L, BIS)$ and $\textit{pound} \defs BUNIT(M, BIS)$. Moreover, we can create derived units such as $foot \defs 1/3 \scaleQ yard$, $inch \defs 1/12 \scaleQ foot$, and $gallon \defs 277.421 \scaleQ inch^3$. Then, we can formally specify that certain quantities are measured according to the BIS.

We can convert quantities between the SI and BIS by the creation of a suitable conversion schema $BSI :: \textit{BIS} \Rightarrow_U \textit{SI}$. The factors for length and mass required for this conversion are 0.9143993 and 0.453592338, respectively. Since time is measured in seconds, and the other dimensions have no interpretation, we set them to 1 in the conversion schema. We can then, for example, convert a BIS quantity of 1 ounce to grams using the conversion $\textit{qconv}_{BSI}(1 \scaleQ ounce) \approx 37.8 \scaleQ gram$\footnote{We use exact rational arithmetic for this in Isabelle/HOL, but we present an approximate decimal expansion for ease of comprehension.}. We also create unit systems for the UCS and CGS systems, with suitable conversion factors.

Whilst we can use quantity conversions between systems directly, it is often more convenient to use the SI as a frame of reference for different unit systems. Indeed, this is a key application of the SI for resolving mismatches between unit systems. We therefore create a type class to representation metrification:

\begin{definition}[Metrifiable Unit Systems]
$$\begin{array}{l}
\mathbf{class}~metrifiable = \textit{unit-system} + \\
\quad  \mathbf{fixes}~convschema :: \mathcal{S}\,itself \Rightarrow (\mathcal{S} \Rightarrow_U SI)
\end{array}$$
\end{definition}

\noindent A unit system $\mathcal{S}$ is metrifiable if there is a conversion schema from $\mathcal{S}$ to \textit{SI}. Consequently, the BIS, UCS, CGS, and the SI itself are all metrifiable. Consequently, for any pair of metrifiable systems, $\mathcal{S}_1$ and $\mathcal{S}_2$, we define a generic conversion function $QMC_{\mathcal{S}_1 \rightarrow \mathcal{S}_2} :: N[\mathcal{D}, \mathcal{S}_1] \Rightarrow N[\mathcal{D}, \mathcal{S}_2]$, which performs conversion via metrification. This function first uses the conversion schema for $\mathcal{S}_1$ to convert to the SI system, and then uses the inverse schema for $\mathcal{S}_2$ to convert from SI to $\mathcal{S}_2$. For example, we can show that $QMC_{CGS \to BIS} (12 \scaleQ centimetre) \approx 4.724 \scaleQ inch$. We can therefore use the Isabelle type system to precisely specify what system a measurement is made in, and seamlessly convert between a variety of other systems.

\section{Conclusions}
\label{sec:conclusions}
In this paper, we have presented a comprehensive mechanisation of the ISQ in Isabelle/HOL. Our mechanisation allows us to precisely define the dimension, unit, and unit system that are employed by a particular system. Moreover, we can use the Isabelle type system to ensure that only measurements of the same dimension and unit system can be combined in a calculation. We have presented a substantial theory development of about 2500 lines of definitions 
and proofs that captures the ISQ and SI as defined in the international standard of the VIM~\cite{bipm-jcgm:2012:VIM}. The theories available on the Isabelle/HOL Archive of Formal Proofs provide a type system for physical quantities and measurements that is by construction sound and  complete. Given the fact that Isabelle's type-system is far from being trivial, we believe that this is both significant and useful for applications in the hybrid system domain. We provided a validation of our theory by checking the mandatory definitions and described corollaries in the VIM and the SI Brochure~\cite{SI-Brochure}. An earlier version of our implementation was also applied in an industrial case study on a formal model for an autonomous underwater vehicle~\cite{foster2020formal}, which provides further validation.

There are a number of directions for future work. The current approach to handling dimension mismatches using coercions could be better automated by using the coercive subtyping mechanism~\cite{Traytel2011CoerciveSubtyping}. This effectively extends the type inference algorithm so that type mismatches can be automatically resolved by insertion of registered coercion functions. At the same time, our approach to characterising dimension types, illustrated in Figure~\ref{fig:dim-mapping}, is not specific to the ISQ, and could be generalised to other problems that are typically solved with dependent types. For example, we could normalise type expressions containing arithmetic operators to relate vectors parametrised by the length. Therefore, in future work we will investigate a generic approach using universe constructions to justify type-level functions and coercions.

\bibliographystyle{IEEEtran}
\bibliography{references}

\end{document}